\documentclass[12pt]{article}
\title{The Stieltjes constants, their relation to the $\eta_j$ coefficients,
and representation of the Hurwitz zeta function}
\author{Mark W. Coffey\\
Department of Physics\\
Colorado School of Mines\\
Golden, CO  80401\\
(Received $\mbox{~~~~~~~~~~~~~~~~~~~~~~~~~~~~~~~2009}$)}
\date{February 18, 2009}  %extension to Prop. 3 added, 3 new refs, some new eqn nos.
\begin{document}
\maketitle
%\vspace{.25cm}
\baselineskip=25 pt
\begin{abstract}
%TBR ...

The Stieltjes constants $\gamma_k(a)$ are the expansion coefficients in
the Laurent series for the Hurwitz zeta function about its only pole at
$s=1$.  We present the relation of $\gamma_k(1)$ to the $\eta_j$
coefficients that appear in the Laurent expansion of the logarithmic 
derivative of the Riemann zeta function about its pole at $s=1$.  
We obtain novel integral representations of the Stieltjes constants
and new decompositions such as $S_2(n) = S_\gamma(n) + S_\Lambda(n)$ for the
crucial oscillatory subsum of the Li criterion for the Riemann hypothesis.
The sum $S_\gamma(n)$ is $O(n)$ and we present various integral 
representations for it.  We present novel series representations of $S_2(n)$.
We additionally present a rapidly convergent expression for $\gamma_k=
\gamma_k(1)$ and a variety of results pertinent to a parameterized
representation of the Riemann and Hurwitz zeta functions.

\end{abstract}
 
\vspace{.25cm}
\baselineskip=15pt
\centerline{\bf Key words and phrases}
\medskip 

\noindent

Stieltjes constants, Riemann zeta function, Hurwitz zeta function, Laurent 
expansion, $\eta_j$ coefficients, von Mangoldt function, Li criterion,
Riemann hypothesis

%\vspace{.25cm}
%\vfill
%\centerline{\bf 2000 PACS numbers}
%TBS ...%02.30.-f, 06.30.Lz     %these are not specific 
 
\baselineskip=25pt
\pagebreak
\medskip
\centerline{\bf Introduction and statement of results}
\medskip

In the Laurent expansion of the Riemann zeta function about $s=1$,
$$\zeta(s)={1 \over {s-1}}+ \sum_{n=0}^\infty {{(-1)^n} \over {n!}}\gamma_n
(s-1)^n, \eqno(1)$$
the Stieltjes constants $\gamma_k$ \cite{coffeyjmaa,coffeyprsa,ivic,kreminski,liang,stieltjes} can be
written in the form
$$\gamma_k=\lim_{N \to \infty}\left (\sum_{m=1}^N {1 \over m}\ln^k m-
{{\ln^{k+1} N} \over {k+1}}\right ). \eqno(2)$$
%and several other forms have been given \cite{scref}.  
From the expansion around $s=1$ of the logarithmic derivative of the
zeta function, 
$${{\zeta'(s)} \over {\zeta(s)}}=-{1 \over {s-1}}-\sum_{p=0}^\infty 
\eta_p (s-1)^p, ~~~~~~|s-1| < 3, \eqno(3)$$ 
we have
$$\ln \zeta(s)=-\ln(s-1)-\sum_{p=1}^\infty {\eta_{p-1} \over p}(s-1)^p.
\eqno(4)$$
The constants $\eta_j$ can be written as
$$\eta_k={(-1)^k \over {k!}} \lim_{N \to \infty}\left (\sum_{m=1}^N {1 \over 
m} \Lambda(m)\ln^k m - {{\ln^{k+1} N} \over {k+1}}\right ), \eqno(5)$$
where $\Lambda$ is the von Mangoldt function 
\cite{edwards,ivic,karatsuba,titch,riemann}, defined by 
$\Lambda(n)=\ln p$ if $n=p^k$ for a prime number $p$ and some integer
$k \geq 1$, and $\Lambda(n)=0$ otherwise.
The radius of convergence of the expansion (3) is $3$, as the first
singularity encountered is the trivial zero of $\zeta(s)$ at $s=-2$.
The Dirichlet series corresponding to Eqs. (3) and (4) valid for Re $s>1$ are
$${{\zeta'(s)} \over {\zeta(s)}}=-\sum_{n=1}^\infty {{\Lambda(n)} \over n^s}
~~~~\mbox{and}~~~~ \ln \zeta(s)=\sum_{n=2}^\infty {{\Lambda(n)} \over {n^s 
\ln n}}.  \eqno(6)$$
Indeed, Eqs. (6) hold on $\sigma=$ Re $s=1$ if $t=$ Im $s \neq 0$ 
\cite{titch}. 

We note that there are various explicit expressions for the von Mangoldt
function.  These include
$$\exp[\Lambda(n)]={{\mbox{LCM}(1,\ldots,n)} \over {\mbox{LCM}(1,\ldots,n-1)}},
\eqno(7)$$
where LCM is the least common multiple function, and Linnik's identity \cite{linnik} (pp. 21-22)
$$\Lambda_1 \equiv {{\Lambda(n)} \over {\ln n}}= -\sum_{k=1}^{\ln n/\ln 2}
{{(-1)^k} \over k} \tau_k'(n).  \eqno(8)$$
In regard to Eq. (8), the strict $\tau_k'$ and exact $\tau_k$ divisor functions
are related by
$$\tau_k'(n)=|\{n_1,\ldots,n_k \geq 2; n_1 \cdots n_k=n\}| \eqno(9a)$$
$$=\sum_{\ell=0}^k (-1)^{k-\ell} {k \choose \ell} \tau_\ell(n).  \eqno(9b)$$
In particular, there is a finite sum on the right side of Eq. (8).

We recall that the function $\ln \zeta(s)$ is intimately connected with the 
prime counting function $\pi(x)$, the number of primes less than $x$.
We have
$$\ln \zeta(s) = s \int_2^\infty {{\pi(x)} \over {x(x^s-1)}} dx.  \eqno(10)$$
Hence the behaviour of the function $\pi(x)$ is related to the important
coefficients $\eta_j$.  For further background on the classical zeta
function we refer to standard texts \cite{edwards,ivic,karatsuba,titch,riemann}.
 
The Hurwitz zeta function, defined by $\zeta(s,a)=\sum_{n=0}^\infty (n+a)^{-s}$
for Re $s>1$ and Re $a>0$ extends to a meromorphic function in the entire
complex $s$-plane.  The generalization of the Laurent expansion (1) is
$$\zeta(s,a)={1 \over {s-1}}+ \sum_{n=0}^\infty {{(-1)^n} \over {n!}}\gamma_n(a)
(s-1)^n. \eqno(11)$$
As shown by Eq. (1), by convention one takes $\gamma_k = \gamma_k(1)$.
%somewhere later to consider the \gamma(1/2) to \gamma(1) relation ...

Having set the notation above we may now state our main results.
{\newline \bf Proposition 1}.  For integers $k \geq 0$ we have 
$${{(-1)^k} \over {k!}}\gamma_k - \eta_k ={{(-1)^k} \over {k!}}
\sum_{m=1}^\infty {{[1-\Lambda(m)]} \over m} \ln^k m, \eqno(12)$$
where the sum on the right starts with $m=2$ when $k>0$.  The case of $k=0$
yields the identity
{\newline \bf Corollary 1}
$$\gamma={1 \over 2}\sum_{m=1}^\infty {{[1-\Lambda(m)]} \over m}, \eqno(13)$$
where $\gamma$ is the Euler constant.

Put
$$S_2(n) \equiv -\sum_{m=1}^n {n \choose m} \eta_{m-1}.  \eqno(14)$$
Let $L_n^\alpha(x)$ be the Laguerre polynomial of degreee $n$ (e.g.,
\cite{andrews,grad}) %or Chihara, Szego, Lebedev, etc. too
and $P_1(t)=B_1(t-[t])=t-[t]-1/2$ the first periodized
Bernoulli polynomial (e.g., \cite{ivic,titch}).  
Then we have the series representation given in
{\newline \bf Proposition 2}.  For integers $n \geq 1$ we have (a)
$$S_2(n) = S_\gamma(n) + S_\Lambda(n), \eqno(15)$$
where
$$S_\gamma(n) \equiv \sum_{k=1}^n {{(-1)^k} \over {(k-1)!}} {n \choose k}
\gamma_{k-1}=\int_1^\infty {1 \over t} L_{n-1}^1(\ln t) dP_1(t), \eqno(16)$$
$$S_\Lambda(n) \equiv \sum_{m=1}^\infty {{[1-\Lambda(m)]} \over m} L_{n-1}^1
(\ln m)$$
$$\equiv n + S_{2\Lambda}(n)=n+\sum_{m=2}^\infty {{[1-\Lambda(m)]} \over m} L_{n-1}^1(\ln m), \eqno(17)$$
(b) $S_\gamma(n) = O(n)$, (c) the average values
$${1 \over M}\sum_{n=1}^M S_\gamma(n) = {1 \over M}\int_1^\infty {1 \over t}
L_{M-1}^2(\ln t) dP_1(t), \eqno(18a)$$
$${1 \over M}\sum_{n=1}^M S_\Lambda(n) = {1 \over M}\sum_{m=1}^\infty
{{[1-\Lambda(m)]} \over m} L_{M-1}^2(\ln m), \eqno(18b)$$
and (d) for $N \geq 1$ a fixed integer 
$$S_\gamma(n)=-\sum_{\nu=1}^N {{L_{n-1}^1(\ln \nu)} \over \nu}-L_n(\ln N)
+1+{1 \over {2N}}L_{n-1}^1(\ln N)+O\left({1 \over {N^{2-\epsilon}}}\right)
L_{n-1}^1(1),$$
where $\epsilon >0$ is arbitrary.

We have found new integral representations of the Stieltjes constants, given in
{\newline \bf Proposition 3}.  Let Re $a>0$.  Then we have (a)
$$\gamma_k(a)= {1 \over {2a}}\ln^k a-{{\ln^{k+1} a} \over {k+1}} + {2 \over a}
\mbox{Re} \int_0^\infty {{(y/a-i)\ln^k (a-iy)} \over {(1+y^2/a^2)(e^{2 \pi y}
-1)}}dy. \eqno(19)$$
(b) Let $n \geq 1$ be an integer.  Then we have
$$\gamma_k(a)=\sum_{m=0}^{n-1} {{\ln^k(m+a)} \over {m+a}}+
{{\ln^k (n+a)} \over {2(n+a)}}-{{\ln^{k+1} (n+a)} \over {k+1}}$$
$$+ {2 \over {n+a}}\mbox{Re} \int_0^\infty {{[y/(n+a)-i]\ln^k (n+a-iy)} \over {[1+y^2/(n+a)^2](e^{2 \pi y}-1)}}dy. \eqno(20)$$
(c)
$$\gamma_m(a)=-{\pi \over 2}{1 \over {m+1}}\mbox{Re}\int_0^\infty {{\ln^{m+1}
(2a-1-it)} \over {\cosh^2 (\pi t/2)}} dt$$
$$+{\pi \over 2}m!\sum_{j=1}^{m+1} {{\ln^j 2} \over {j!}} {{(-1)^{j+1}} \over
{(m-j+1)!}} \mbox{Re} \int_0^\infty {{\ln^{m-j+1}(2a-1-it)} \over {\cosh^2 
(\pi t/2)}}dt.  \eqno(21)$$
Let $\psi(z)=\Gamma'(z)/\Gamma(z)$ be the digamma function, where $\Gamma$ is
the Gamma function \cite{nbs,andrews,grad}.  From Proposition 3 follows
Corollaries 2 and 3.
%{\bf Corollaries 2}--this for both $S_\gamma$ and $\gamma_0(a)$
From the case $m=0$ in Proposition 3 we have
{\newline \bf Corollary 2}. (a)
$$\gamma_0(a)=-\psi(a)={1 \over {2a}}-\ln a + {2 \over a}
\mbox{Re} \int_0^\infty {{(y/a-i)} \over {(1+y^2/a^2)(e^{2 \pi y}
-1)}}dy, \eqno(22)$$
(b) for integers $n \geq 1$,
$$\gamma_0(a)=-\psi(a)=\sum_{m=0}^{n-1}{1 \over {m+a}}+{1 \over {2(n+a)}}
-\ln (n+a) + {2 \over {(n+a)}}\mbox{Re} \int_0^\infty {{[y/(n+a)-i]} \over {[1+y^2/(n+a)^2](e^{2 \pi y}-1)}}dy, \eqno(23)$$
and (c)
$$\gamma_0(a)=-\psi(a)=-{\pi \over 2}\mbox{Re}\int_0^\infty {{\ln(2a-1-it)}
\over {\cosh^2 (\pi t/2)}} dt + \ln 2$$
$$=-{\pi \over 4}\int_0^\infty {{\ln[(2a-1)^2+t^2]} \over {\cosh^2(\pi t/2)}} dt
+ \ln 2.  \eqno(24)$$

%Next is Corollary 3 on $S_\gamma(n)$ from Proposition 3 ..... TO DO .....
Based upon the $a=1$ case of Proposition 3 we obtain additional new integral
representations of the sum $S_\gamma$ defined in Eq. (16):
{\newline \bf Corollary 3}.
$$S_\gamma(n)=-\int_0^\infty {1 \over {(1+y^2)(e^{2\pi y}-1)}}\left[(y-i)
L_{n-1}^1[\ln(1-iy)]+(y+i)L_{n-1}^1[\ln(1+iy)]\right] dy, \eqno(25)$$
and
$$S_\gamma(n)={\pi \over 4}\int_0^\infty [L_n(2a-1-it)+L_n(2a-1+it)-2] {{dt}
\over {\cosh^2(\pi t/2)}}$$
$$-{\pi \over 4}\sum_{j=1}^n {{\ln^j 2} \over {j!}} {n \choose j}\int_0^\infty
\{_1F_1[j-n;j+1;\ln(2a-1-it)]+ ~_1F_1[j-n;j+1;\ln(2a-1+it)]\} {{dt}
\over {\cosh^2(\pi t/2)}},  \eqno(26)$$
where $_1F_1$ is the confluent hypergeometric function, with ${n \choose j} ~_1F_1(j-n;j+1;x)=L_{n-j}^j(x)$.

Let $d(n)$ be the number of divisors of $n$.  Then we have
{\newline \bf Proposition 4}.  (a)  We have for integers $m \geq 0$
$$(-1)^m \sum_{n=1}^\infty {{\ln^m n} \over n}[d(n)-\ln n -2\gamma]=(-1)^{m+1}
\left(1+{2 \over {m+1}}\right)\gamma_{m+1}-2\gamma (-1)^m \gamma_m$$
$$+\sum_{\ell=0}^m (-1)^\ell {m \choose \ell} \gamma_\ell \gamma_{m-\ell}.
\eqno(27)$$
In particular we have at $m=0$
{\newline \bf Corollary 4}. 
$$\sum_{n=1}^\infty {1 \over n}[d(n)-\ln n -2\gamma]=-3\gamma_1-\gamma^2
=-3\gamma_1-{1 \over 4} \left(\sum_{m=1}^\infty {{[1-\Lambda(m)]} \over m}
\right)^2, \eqno(28)$$
giving
{\newline \bf Corollary 4'}. 
$$\sum_{m=1}^\infty {{d(m)} \over m}-\lim_{n\to \infty}\left[{1 \over 2}\ln^2 n
+2\gamma \ln n\right] = \gamma^2-2\gamma_1 > 0.$$
(b) We have for integers $m \geq 0$
$$(-1)^m \sum_{n=1}^\infty {{\ln^m n} \over n}[d(n)-\Lambda(n) \ln n -2\gamma]=(m+1)\eta_{m+1} + {{2(-1)^{m+1}} \over {m+1}}\gamma_{m+1}
-2\gamma (-1)^m \gamma_m$$
$$+\sum_{\ell=0}^m (-1)^\ell {m \choose \ell} \gamma_\ell \gamma_{m-\ell}.
\eqno(29)$$
In particular we have at $m=0$
{\newline \bf Corollary 5}. 
$$\sum_{n=1}^\infty {1 \over n}[d(n)-\Lambda(n)\ln n -2\gamma]=\eta_1-2\gamma_1
- \gamma^2=0.  \eqno(30)$$
We may obtain new series representations of $S_\gamma(n)$ from part (a) and of
$S_2(n)$ from part (b).  As an illustration we have
{\newline \bf Corollary 6}.  For integers $n \geq 1$ we have
$$S_2(n)=\gamma n+\sum_{m=2}^\infty {{L_{n-1}^1(\ln m)} \over {m \ln m}}
[d(m)-\Lambda(m)\ln m-2\gamma]-2\sum_{m=2}^n (-1)^m {n \choose m}
{\gamma_{m-1} \over {(m-1)(m-1)!}}$$
$$-2\gamma\sum_{m=2}^n (-1)^m {n \choose m} {\gamma_{m-2} \over {(m-1)!}}
+\sum_{m=2}^n {n \choose m}{1 \over {(m-1)!}} \sum_{\ell=0}^{m-2} (-1)^\ell
{{m-1} \choose \ell} \gamma_\ell \gamma_{m-\ell-2}.  \eqno(31)$$

Let $\mu(n)$ be the M\"{o}bius function.  
{\newline \bf Proposition 5}.  Then we have (i)
$$\eta_0=-\gamma={1 \over 2}\sum_{k=2}^\infty {{\mu(k)} \over k} \ln^2 k,
\eqno(32)$$
and for integers $\ell \geq 1$
$$(-1)^\ell \eta_\ell={1 \over {(\ell+2)!}}\sum_{k=2}^\infty {{\mu(k)} \over k} \ln^{\ell+2} k + \sum_{k=2}^\infty {{\mu(k)} \over k}\sum_{p=1}^\ell {{\ln^p k}
\over {p!}}{\gamma_{\ell-p+1} \over {(\ell-p)!}}.  \eqno(33)$$
(ii) For integers $j \geq 1$ we have
$$\eta_j={{(-1)^{j+1}} \over {(j+1)!}}\sum_{\ell=2}^\infty {{\mu(\ell)} \over
\ell} \ln^{j+2} \ell+ {{(-1)^j} \over {j!}} \gamma_j-\sum_{\ell=2}^\infty
{{\mu(\ell)} \over \ell} \sum_{n=0}^{j-1} {{(-1)^n} \over {n!}} \gamma_n
{{(-1)^j} \over {(j-n)!}} \ln^{j-n+1} \ell .  $$ %add an \eqno later ... ...
From Proposition 5(i) follows
{\newline \bf Corollary 7}.  We have
$$S_2(n)=\sum_{k=2}^\infty {{\mu(k)} \over k}\ln k\left[{1 \over {n+1}}L_n^1(
\ln k) -1\right]+\sum_{k=2}^\infty {{\mu(k)} \over k} \sum_{\ell=1}^n (-1)^\ell
{n \choose \ell} \sum_{p=0}^{\ell-1} {{\ln^p k} \over {p!}}{\gamma_{\ell-p} 
\over {(\ell-p-1)!}}.  \eqno(34)$$

{\bf Proposition 6}.  Let $B_j$ represent the Bernoulli numbers.
For integers $m \geq 1$ and Re $\lambda > 0$ we have 
$$\gamma_m=-{{m!} \over {1+\lambda}}\sum_{\ell=1}^m {{B_{m-\ell+1}} \over {(m-\ell+1)!}}{{\ln^{m-\ell}2} \over {\ell!}}\sum_{k=1}^\infty \left({\lambda 
\over {1+\lambda}}\right)^k \sum_{j=1}^k (-1)^j {k \choose j} {1 \over \lambda^j} {{\ln^\ell (j+1)} \over {(j+1)}}$$
$$-{1 \over {(1+\lambda)\ln 2}}{1 \over {(m+1)}}\sum_{k=1}^\infty \left({\lambda \over {1+\lambda}}\right)^k \sum_{j=1}^k (-1)^j {k \choose j} {{\ln^{m+1} (j+1)} \over {(j+1)}}-{B_{m+1} \over {(m+1)}} \ln^{m+1} 2. \eqno(35)$$
For $m=0$ we have
$$\gamma_0=\gamma={1 \over 2}\ln 2-{1 \over {(1+\lambda)\ln 2}}\sum_{k=1}^\infty \left({\lambda \over {1+\lambda}}\right)^k \sum_{j=1}^k (-1)^j {k \choose j} {{\ln (j+1)} \over {(j+1)}}.  \eqno(36)$$

{\bf Proposition 7}.  (parameterized series representation of the Hurwitz zeta
function)
%Let $\zeta(s,a)=\sum_{n=0}^\infty (n+a)^{-s}$, Re $s>0$,
%be the Hurwitz zeta function.  Then
For Re $s >1$, Re $a >0$, and Re $\lambda>0$
we have the representation
$$\zeta(s,a)=\sum_{k=0}^\infty {\lambda^k \over {(1+\lambda)^{k+1}}}\sum_{j=0}^k
{k \choose j} {1 \over \lambda^j} {1 \over {(a+j)^s}}.  \eqno(37)$$

{\bf Proposition 8}.  Let the polylogarithm function Li$_s(z)=\sum_{k=1}^\infty z^k/k^s=z\Phi(z,s,1)$ where $\Phi$ is the Lerch zeta function, $s \in C$ when
$|z|<1$ and Re $s >1$ when $|z| =1$.  Then for 
$q \geq 1$, Re $t >0$, and $m > 0$ an integer we have the integral
$$\int_0^1 {u^{t-1} \over {\ln u}}(1-u)^{m-1} \mbox{Li}_q(1-u) du
=\sum_{n=m}^\infty {1 \over {(n-m+1)^q}} \sum_{j=0}^n (-1)^j {n \choose j}
\ln(t+j),  \eqno(38)$$
giving
{\newline \bf Corollary 8}.  At $q=1$, with Li$_1(z)=-\ln(1-z)$ we have the
special case
$$-B(t,m)=\sum_{n=m}^\infty {1 \over {(n-m+1)}} \sum_{j=0}^n (-1)^j {n \choose j}
\ln(t+j),  \eqno(39)$$
where $B(x,y)=\Gamma(x)\Gamma(y)/\Gamma(x+y)$ is the Beta function \cite{andrews}.
%and $\Gamma$ is the Gamma function \cite{andrews}.

Let for Re $x>0$ and Re $y > 0$
$$B(z_1,z_2,x,y) \equiv \int_{z_1}^{z_2} t^{x-1}(1-t)^{y-1} dt \eqno(40)$$
be the generalized incomplete Beta function and Ei$(x)\equiv \int_{-\infty}^x
(e^t/t) dt$ the exponential integral.  Then we have 
{\newline \bf Proposition 9} (context of the logarithmic integral)
$$\int_{z_1}^{z_2} {{(u^{t-1}-u^{a-1})} \over {\ln u}} (1-u)^{y-1} du
=\sum_{j=0}^\infty (-1)^j {{y-1} \choose j} \left\{\mbox{Ei}[(a+j)\ln z_1]
-\mbox{Ei}[(t+j)\ln z_1] \right.$$
$$\left. -\mbox{Ei}[(a+j)\ln z_2]+\mbox{Ei}[(t+j)\ln z_2]\right \}, \eqno(41)$$
giving
{\newline \bf Corollary 9}
$$\int_{z_1}^{z_2} {{(u^{t-1}-u^{a-1})} \over {\ln u}} du
=\mbox{Ei}[a\ln z_1]-\mbox{Ei}[t\ln z_1]-\mbox{Ei}[a\ln z_2]+\mbox{Ei}[t)\ln z_2]$$
$$-\mbox{Ei}[(a+1)\ln z_1]+\mbox{Ei}[(t+1)\ln z_1]+\mbox{Ei}[(a+1)\ln z_2]-\mbox{Ei}[(t+1)\ln z_2].  \eqno(42)$$

Propositions 1 and 2 were first obtained by the author several years ago as complements to Refs. \cite{coffeyjcam04} and \cite{coffeympag} concerning the Li
criterion for the Riemann hypothesis.  The proof of Proposition 2 given in the 
sequel is more direct than the original. The sort of alternating binomial sums
that occur in Proposition 6 motivates a study of integrals such as appear in Propositions 8 and 9. After the proofs of these Propositions we present Discussion that contains additional examples and extensions.

\medskip
%\pagebreak
\centerline{\bf Proofs of the Propositions}
\medskip

{\em Proposition 1}.  In order to derive Eq. (12), we add Eqs. (1) and (3) and
make use of Eq. (6), resulting in
$$\zeta(s)+{{\zeta'(s)} \over {\zeta(s)}}=\sum_{n=0}^\infty \left[{{(-1)^n} 
\over {n!}}\gamma_n - \eta_n\right ](s-1)^n=\sum_{k=1}^\infty {{[1-\Lambda(k)]}
\over {k^s}},  \eqno(43)$$
wherein the pole at $s=1$ has been removed.  Repeated differentiation of Eq.
(43) gives
\newline{\bf Corollary 10}
$$\zeta^{(j)}(s)+\left[{{\zeta'(s)} \over {\zeta(s)}}\right]^{(j)}=
\sum_{n=j}^\infty \left[{{(-1)^n} \over {n!}}\gamma_n - \eta_n\right ]
n(n-1)(n-2) \cdots (n-j+1)(s-1)^{n-j}$$
$$=(-1)^j \sum_{k=1}^\infty {{[1-\Lambda(k)]}\over {k^s}}\ln^j k,  \eqno(44)$$
where the sum on the right starts with $k=2$ when $j>0$.
Taking $s \to 1^+$ in Eq. (44) gives Eq. (12).

Remarks.  (i) Taking different values of $s$ in Eq. (44) yields various
connections between the $\eta$'s and the Stieltjes constants.  For instance,
with $s=2$ in Eq. (44) we have
$$\sum_{n=j}^\infty \left[{{(-1)^n} \over {n!}}\gamma_n - \eta_n\right ]
(-n)_j=\sum_{k=1}^\infty {{[1-\Lambda(k)]}\over {k^2}}\ln^j k,  \eqno(45)$$
where $(a)_\ell=\Gamma(a+\ell)/\Gamma(a)$ is the Pochhammer symbol.  
(ii)  For Eq. (13) we use the values $\gamma_0=\gamma$ and $\eta_0=-\gamma$
(e.g., \cite{coffeyjcam04,coffeympag}).
This special case (13) for the Euler constant has unfortunately appeared in
several places in the literature including \cite{gourdon} and \cite{havil}
(p. 109) with the sum starting at $2$ instead of $1$, missing a dominant
contribution of $1/2$.  (iii)  Proposition 1 may also be proved by forming
$$\ln \zeta(s)-\int_s^\infty [\zeta(y)-1]dy=\sum_{n=2}^\infty {{[\Lambda(n)
-1]} \over {n^s \ln n}}$$
$$=\gamma-C_{1,1}+\sum_{j=1}^\infty\left[-{\eta_{j-1} \over j} + {{(-1)^{j+1}} \over {j!}} \gamma_{j-1}\right](s-1)^j, \eqno(46)$$
where \cite{boas79} (p. 156)
$$C_{1,1} \equiv \lim_{x \to \infty}\left [\sum_{n=2}^x {1 \over
{n \ln n}} - \ln(\ln x)\right ] \simeq 0.7946786.$$
The one new sum that appears here has value
$$\sum_{n=2}^\infty{{[\Lambda(n)-1]} \over {n \ln n}} = \gamma - C_{1,1}
\simeq -0.217464.$$
%have to check this ..................  
(iv)  The Stieltjes and $\eta_j$ constants are strongly connected and one may
easily write a recursion relation between the two sets of constants 
\cite{coffeyjcam04} (Appendix).  For instance, $\eta_1=\gamma^2+2\gamma_1$.
Then from Eq. (12) we may write either
$$\eta_1={1 \over 3}\gamma^2 + {2 \over 3}\sum_{m=1}^\infty {{[1-\Lambda(m)]}
\over m} \ln m, \eqno(47)$$
or 
$$\gamma_1={1 \over 3}\left[\sum_{m=1}^\infty {{[1-\Lambda(m)]}\over m} \ln m
-\gamma^2\right],$$
with the corresponding formula for $\gamma$ given in Eq. (13).  Similarly, we
have corollary expressions for all of the $\gamma_k$ and $\eta_j$ constants.

{\em Proposition 2}.  
By applying Proposition 1, the definition (14) of $S_2(n)$, and the power series
definition of $L_n^\alpha$, we obtain
$$S_2(n)=\sum_{k=1}^n {{(-1)^k} \over {(k-1)!}} {n \choose k} \gamma_{k-1}
+S_\Lambda(n).  \eqno(48)$$
The second line of Eq. (17) follows since $\Lambda(1)=0$ and $L_{n-1}^1(0)=n$.
We obtain alternative forms of the sum $S_\gamma(n)$ of Eq. (16) by using
the integral representation \cite{ivic}
$$\gamma_k=-\int_1^\infty {1 \over t} \ln^k t dP_1(t)
=\int_1^\infty {{\ln^{k-1} t} \over t^2}(k- \ln t) P_1(t)dt -\delta_{k0}/2
\eqno(49)$$
$$={{(-1)^{k-1}} \over {(k-1)!}}\int_1^\infty P_k(t)\left({d \over {dt}}\right)^k {{\ln^k t} \over t}dt - \delta_{k0}/2,  \eqno(50)$$
where $\delta_{jk}$ is the Kronecker symbol.
%NEED TO CHECK factorial factor in Eq. (50), done ...
In Eq. (50), where we integrated by parts $k-1$ times, $P_k(t)=B_k(t-[t])$ with
$B_k$ the $k$th degree Bernoulli polynomial, such that $P_{k+1}'(t)=(k+1)P_k(t)$.   
%see p. 1078 of G/R
By then applying the definition (16) and the power series form of the associated
Laguerre polynomials we have
$$S_\gamma(n)=\int_1^\infty {{P_1(t)} \over t^2}\left[L_{n-2}^2(\ln t)
+ L_{n-1}^1(\ln t) \right] dt + n/2.  \eqno(51)$$
This representation of $S_\gamma$ is equivalent to integration by parts on
the expression given on the right side of Eq. (16).
%TBS/finish later today .....................

For part (b) we observe $P_1(t)=O(1)$.  So for a constant $C >0$ the integral
term in Eq. (51) is majorized by
$$C\int_1^\infty {1 \over t^2}\left[L_{n-2}^2(\ln t) + L_{n-1}^1(\ln t) \right]
dt = C\int_0^\infty e^{-u}\left[L_{n-2}^2(u)+ L_{n-1}^1(u) \right] du$$
$$=C\left(1 + {1 \over {n-2}}\right), ~~~~~~~~n \geq 3,  \eqno(52)$$
where we used the value of a Laplace transform \cite{grad}
$$\int_0^\infty e^{-x} L_{n-\nu}^\nu (x) dx = {1 \over {\Gamma(\nu)}}
{{(n-1)!} \over {(n-\nu)!}}.  \eqno(53)$$
Then $S_\gamma(n)=O(n)$.  For part (c) we apply Eqs. (16) and (17) and 
interchange the order of operations. % and Proposition 2 is completed.  
%is S_2(n) possibly
%larger, not smaller as $O(n^{3/4}$? to consider ...
%comment in the Discussion section on the implications of $S_\Lambda$'s 
%order for the validity of the RH, in the context of the Li criterion ...

For part (d), we make use of \cite{liang} (Section 2)
$$\gamma_k=\sum_{\nu=1}^n {{\ln^k \nu} \over \nu}-{{\ln^{k+1} n} \over {k+1}}
-{1 \over 2} {{\ln^k n} \over n} + O\left({1 \over n^{2-\epsilon}}\right), \eqno(54)$$
with $\epsilon >0$ arbitrary.  This equation is essentially an asymptotic
form of $\gamma_k$ for $k \gg 1$.  We substitute this equation into the
definition (16) of $S_\gamma(n)$, apply the power series form of the
associated Laguerre polynomials, and Proposition 2 is completed.

{\em Proposition 3}.  Part (a) is based upon the well known Hermite formula
for $\zeta(s,a)$:
$$\zeta(s,a)={a^{-s} \over 2} + {a^{1-s} \over {s-1}} + 2 \int_0^\infty 
{{\sin(s \tan^{-1} y/a)} \over {(y^2+a^2)^{s/2}}} {{dy} \over {(e^{2\pi y}-1)}}.  \eqno(55)$$
We use the expression 
$$\tan^{-1} w = {1 \over {2i}} \ln\left({{1+iw} \over {1-iw}}\right ), \eqno(56)$$ 
so that
$\exp[\pm is \tan^{-1} (y/a)]= [(1+iy/a)/(1-iy/a)]^{\pm s/2}$.  Then the
integral term in Eq. (55) becomes
$$2 \int_0^\infty {{\sin(s \tan^{-1} y/a)} \over {(y^2+a^2)^{s/2}}} {{dy} 
\over {(e^{2\pi y}-1)}}~~~~~~~~~~~~~~~~~~~~~~~~~~~~~~~~~~~~~~~~~~~~~~~~~~~~~~~~~~~$$
$$=-i\int_0^\infty{1 \over {(y^2+a^2)^{s/2}}}\left[ \left(
{{1+iy/a} \over {1-iy/a}}\right)^{s/2}- \left({{1-iy/a} \over {1+iy/a}}\right)^{s/2}
\right] {{dy} \over {(e^{2\pi y}-1)}}$$
% =eqn at bottom of p. 1 of my 9/21/06 notes to insert next ....  
$$={-i \over a^s}\int_0^\infty {1 \over {(1+y^2/a^2)}} \left[{{(1+iy/a)} \over
{(1-iy/a)^{s-1}}} - {{(1-iy/a)} \over {(1+iy/a)^{s-1}}} \right]{{dy} \over
{(e^{2 \pi y} -1)}}$$
$$={1 \over a}\int_0^\infty {1 \over {(1+y^2/a^2)}}\left[(y/a-i)e^{-(s-1)\ln(a-iy)}
+(y/a+i)e^{-(s-1)\ln(a+iy)}\right] {{dy} \over {(e^{2 \pi y} -1)}}.  \eqno(57)$$
For the other terms in Eq. (55) we have
$${a^{-s} \over 2} + {a^{1-s} \over {s-1}}={{e^{-(s-1)\ln a}} \over {2a}}
+ {{e^{-(s-1)\ln a}} \over {s-1}}$$
$$={1 \over {s-1}} + {1 \over {2a}}\sum_{j=0}^\infty {{(-1)^j} \over {j!}} \ln^j a
(s-1)^j+\sum_{j=0}^\infty {{(-1)^{j+1}} \over {(j+1)!}} \ln^{j+1} a (s-1)^j.
\eqno(58)$$
We then expand the right side of Eq. (57) in powers of $s-1$, combine the
result with Eq. (58), and compare with the defining expansion (11) for $\gamma_k(a)$.  We find
$$\gamma_k(a)={1 \over {2a}}\ln^k a-{{\ln^{k+1} a} \over {k+1}} + {1 \over a}
\int_0^\infty {{(y/a-i)\ln^k (a-iy)+(y/a+i)\ln^k (a+iy)} \over {(1+y^2/a^2)
(e^{2 \pi y}-1)}}dy, \eqno(59)$$
that is the same as Eq. (19) and part (a) is proved.

For part (b) we apply Abel-Plana summation (e.g., \cite{sri} p. 90) to write for all
complex $s \neq 1$
$$\zeta(s,a)=\sum_{k=0}^{n-1}{1 \over {(k+a)^s}}+{(n+a)^{-s} \over 2} + {(n+a)^{1-s} \over {s-1}} + 2 \int_0^\infty {{\sin[s \tan^{-1} y/(n+a)]} \over {[y^2+(n+a)^2]^{s/2}}} {{dy} \over {(e^{2\pi y}-1)}}.  \eqno(60)$$
We then expand the terms of this equation in powers of $s-1$ in like manner to
part (a).

Part (c) makes use of the integral representation for Re $a > 1/2$ \cite{jensen}
$$\zeta(s,a)={{\pi 2^{s-2}} \over {(s-1)}}\int_0^\infty [t^2+(2a-1)^2]^{(1-s)/2}
{{\cos[(s-1)\tan^{-1}[t/(2a-1)]} \over {\cosh^2 (\pi t/2)}}dt.  \eqno(61)$$
We again apply Eq. (56) so that we are able to write
$$\zeta(s,a)={\pi \over 4}{1 \over {s-1}}\sum_{j=0}^\infty {{\ln^j 2} \over {j!}}
(s-1)^j \int_0^\infty \left[{1 \over {(2a-1-it)^{s-1}}}+{1 \over {(2a-1+it)^{s-1}}}
\right] {{dt} \over {\cosh^2 (\pi t/2)}}$$
$$={\pi \over 4}{1 \over {s-1}}\int_0^\infty \sum_{\ell=0}^\infty {{(-1)^\ell} \over {\ell !}}\left [\ln^\ell (2a-1-it) + \ln^\ell (2a-1+it)\right] (s-1)^\ell {{dt} \over {\cosh^2 (\pi t/2)}}$$
$$+{\pi \over 4}{1 \over {s-1}}\sum_{j=1}^\infty {{\ln^j 2} \over {j!}}
\sum_{\ell=0}^\infty {{(-1)^\ell} \over {\ell !}} \int_0^\infty \left[
\ln^\ell (2a-1-it) + \ln^\ell (2a-1+it)\right] (s-1)^{j+\ell}{{dt} \over {\cosh^2 (\pi t/2)}}. \eqno(62)$$
We then (i) separate out the $\ell=0$ term two lines above and use the integral
$\int_0^\infty \mbox{sech}^2(\pi t/2) dt = 2/\pi$, and (ii) reorder the double sum in the last line of Eq. (57), thereby obtaining
$$\zeta(s,a)={1 \over {s-1}}+{\pi \over 4} \sum_{\ell=0}^\infty {{(-1)^{\ell+1}} \over {(\ell+1)!}}\int_0^\infty \left [\ln^{\ell+1} (2a-1-it) + \ln^{\ell+1} (2a-1+it)\right] {{dt} \over {\cosh^2 (\pi t/2)}} (s-1)^\ell$$
$$+{\pi \over 4}\sum_{m=0}^\infty (-1)^m \sum_{j=1}^{m+1} {{\ln^j 2} \over {j!}}
{{(-1)^{j+1}} \over {(m-j+1)!}} \int_0^\infty \left[\ln^{m-j+1} (2a-1-it) + 
\ln^{m-j+1} (2a-1+it)\right]$$
$$~~~~~~~~~~~~~~~~~~~~~~~~~~~~~~~~~~\times {{dt} \over {\cosh^2 (\pi t/2)}} (s-1)^m. \eqno(63)$$
Comparing Eq. (63) with the Laurent expansion (11) gives part (b) and 
Proposition 3 is proved.

For Corollary 2 we put $m=0$ in Eqs. (19), (20), and (21) and use the fact that
$\gamma_0(x) = -\psi(x)$ (e.g., \cite{coffeyjmaa}).  

Remarks.  As it must, Eq. (60) satisfies $\zeta(0,a)=1/2-a$, $\zeta'(0,a)=\ln 
\Gamma(a)-(1/2)\ln(2\pi)$, and $B_m(x)=-m\zeta(1-m,x)$ for positive integers $m$.

Equation (22) recovers the result of differentiating Binet's second
expression for $\ln \Gamma(z)$.  Equation (24) subsumes the special case at
$a=1/2$ given in \cite{grad} (p. 580).

For Corollary 3 we apply the definition (16) of $S_\gamma(n)$ together with
{\newline \bf Lemma 1}.  We have
$$\sum_{j=\nu}^n {{(-1)^{j-1}} \over {(j-\nu)!}}{n \choose j} w^{j-\nu}
=(-1)^{\nu-1} L_{n-\nu}^\nu (w),  \eqno(64)$$
that follows from the power series form of the associated Laguerre polynomials.
The second line of Eq. (26) follows first from reordering a double sum to obtain
$$\sum_{k=1}^n (-1)^k {n \choose k} \sum_{j=1}^k {{\ln^j 2} \over {j!}} 
{{(-1)^{j+1}} \over {(k-j)!}} \ln^{k-j} (2a-1 \mp it)$$
$$=\sum_{j=1}^n \sum_{k=0}^{n-j} (-1)^{k+j} {n \choose {k+j}}{{\ln^k(2a-1 \mp it)}
\over {k!}} {{\ln^j 2} \over {j!}} (-1)^{j+1}.  \eqno(65)$$
We then use various relations to express ${n \choose {k+j}}$ in terms of
Pochhammer symbols:
$${n \choose {k+j}}={{(-1)^{k+j}} \over {(k+j)!}} (-n)_{j+k}=
{{(-1)^{k+j}} \over {(k+j)!}}(-n)_j (j-n)_k=
{{(-1)^{k+j}} \over {(k+j)!}}{{(-1)^j n!} \over {(n-j)!}}(j-n)_k, \eqno(66)$$
%add an eqno in here later .............................................. DONE
where $(k+j)!=j! (j+1)_k$.  We then apply the power series definition of
the confluent hypergeometric function, Eq. (26) follows and Corollary 3 is
completed.  

Remarks.  Similar to Proposition 3 we have obtained further integral
representations of the Stieltjes constants and the sum $S_\gamma(n)$ from
related integral representations of the Riemann zeta function.  In particular,
we have \cite{jensen}
$$\zeta(s)={2^{s-1} \over {s-1}} - 2^s \int_0^\infty (1+t^2)^{-s/2} \sin(s
\tan^{-1} t) {{dt} \over {e^{\pi t} +1}}, \eqno(67)$$ %add eqno later .....
and
$$\zeta(s)={2^{s-1} \over {1-2^{1-s}}}\int_0^\infty (1+t^2)^{-s/2} {{\cos(
s \tan^{-1}t)} \over {\cosh (\pi t/2)}} dt. \eqno(68)$$
As a byproduct we have from Eq. (68)
{\newline \bf Corollary 11}.  For all complex $s \neq 1$ we have
$$\zeta(s)={1 \over {2(1-2^{1-s})}}\int_0^\infty \left[{1 \over {(1/2-iw)^s}}
+ {1 \over {(1/2+iw)^s}}\right ] {{dw} \over {\cosh(\pi w)}}$$
$$={1 \over {2(1-2^{1-s})}}\int_{-\infty}^\infty {1 \over {(1/2-iw)^s}}
{{dw} \over {\cosh(\pi w)}}. \eqno(69)$$
Equation (69) follows from Eq. (68) with a simple change of variable and
the use of relation (56).  

Of a slightly different flavor, we have used the representation \cite{debruijn}
valid for $0 < \mbox{Re} ~s < 2$ 
$$\zeta(s) = {1 \over {s-1}}+{{\sin \pi s} \over {\pi(s-1)}}\int_0^\infty
\left[\psi'(t) - {1 \over {1+t}}\right ] t^{s-1} dt, \eqno(70)$$
to obtain
{\newline \bf Corollary 12}.  We have for integers $m \geq 0$
$$\gamma_m=(-1)^m m!\sum_{k=0}^{[m/2]} {{(-1)^{k+1}} \over {(2k+1)!}}{\pi^{2k}
\over {(m-2k)!}} \int_0^\infty \ln^{m-2k} t \left[\psi'(t) - {1 \over {1+t}}
\right ] dt. \eqno(71)$$
The proof of Corollary 12 makes use of the Taylor series 
$$\sin \pi s = \sum_{k=0}^\infty {{(-1)^{k+1}} \over {(2k+1)!}} \pi^{2k+1}
(s-1)^{2k+1}, \eqno(72)$$
together with series manipulations.  The case of $m=0$ in Eq. (71) 
recovers $\gamma_0=\gamma$, as is readily seen by a limiting argument.
Corollary 12 provides another form of the sum $S_\gamma(n)$, that we omit.

We have found that Proposition 3(a) subsumes the $a=1$ case derived in
Ref. \cite{ainsworth} by a contour integration.  This reference evidences
that Proposition 3 should be a practical method for computing $\gamma_k(a)$.
Indeed, there is now an arbitrary precision Python implementation \cite{mpmath}.

{\em Proposition 4}.  For part (a) we form the combination of Dirichlet series
$$\zeta^2(s)+\zeta'(s)-2\gamma \zeta(s)=\sum_{n=1}^\infty {1 \over n^s}
[d(n)- \ln n-2\gamma], \eqno(73)$$
wherein simple and double polar terms have been eliminated.  By the use of
Eq. (1) and series manipulations we find 
$$\sum_{n=1}^\infty {1 \over n^s} [d(n)- \ln n-2\gamma]~~~~~~~~~~~~~~~~~~~
~~~~~~~~~~~~~~~~~~~~~~~~~~~~~~~~~~~~~~~~~~~~~~~~~~~$$
$$=\sum_{j=0}^\infty
\left [{{2(-1)^{j+1}} \over {(j+1)!}}\gamma_{j+1}+{{(-1)^{j+1}} \over {j!}}
\gamma_{j+1}-2\gamma {{(-1)^j} \over {j!}}\gamma_j+\sum_{\ell=0}^j {{(-1)^\ell}
\over {\ell! (j-\ell)!}} \gamma_\ell \gamma_{j-\ell}\right](s-1)^j.
\eqno(74)$$
Taking the limit as $s \to 1^+$ in this equation gives Corollary 4.  
Corollary 4' then follows from the limit relations (2).  Taking $m$ 
derivatives with respect to $s$ in Eq. (74) and putting $s \to 1^+$ yields
$$(-1)^m\sum_{n=1}^\infty {{\ln^m n} \over n} [d(n)- \ln n-2\gamma]~~~~~~~~~~~~~~
~~~~~~~~~~~~~~~~~~~~~~~~~~~~~~~~~~~~~~~~~~~~~~~~~~~~~~~~~~~$$
$$=\left[
{{(-1)^{m+1}} \over {m!}}\left(1+{2 \over {m+1}}\right)\gamma_{m+1}-2\gamma
{{(-1)^m} \over {m!}} \gamma_m+\sum_{\ell=0}^m {{(-1)^\ell} \over {\ell!
(m-\ell)!}} \gamma_\ell \gamma_{m-\ell}\right ] m!, \eqno(75)$$
and this gives Eq. (27).

For part (b) we form the combination of Dirichlet series
$$\zeta^2(s)-\left({{\zeta'} \over  \zeta}\right)'(s)-2\gamma \zeta(s) = \sum_{n=1}^\infty {1 \over n^s} [d(n)- \Lambda(n)\ln n-2\gamma]. \eqno(76)$$
We then use both Eqs. (1) and (3) and series manipulations to find
$$\sum_{n=1}^\infty {1 \over n^s} [d(n)- \Lambda(n)\ln n-2\gamma]
~~~~~~~~~~~~~~~~~~~~~~~~~~~~~~~~~~~~~~~~~~~~~~~~~~~~~~~~~~~~~~~~~~~~~$$
$$=\sum_{j=0}^\infty
\left [{{2(-1)^{j+1}} \over {(j+1)!}}\gamma_{j+1}+(j+1)\eta_{j+1}
-2\gamma {{(-1)^j} \over {j!}}\gamma_j+\sum_{\ell=0}^j {{(-1)^\ell}
\over {\ell! (j-\ell)!}} \gamma_\ell \gamma_{j-\ell}\right](s-1)^j,
~~~~~~ |s-1| < 3.  \eqno(77)$$
Taking the limit as $s \to 1^+$ in this equation gives Corollary 5.  More 
generally, by taking $m$ derivatives of relation (77) we have
$$(-1)^m\sum_{n=1}^\infty {{\ln^m n} \over n^s} [d(n)- \Lambda(n)\ln n-2\gamma]
~~~~~~~~~~~~~~~~~~~~~~~~~~~~~~~~~~~~~~~~~~~~~~~~~~~~~~~~~~~~~~~~~~~~~~$$
$$=\sum_{j=m}^\infty
\left [{{2(-1)^{j+1}} \over {(j+1)!}}\gamma_{j+1}+(j+1)\eta_{j+1}
-2\gamma {{(-1)^j} \over {j!}}\gamma_j+\sum_{\ell=0}^j {{(-1)^\ell}
\over {\ell! (j-\ell)!}} \gamma_\ell \gamma_{j-\ell}\right]$$
$$~~~~~~~~~~~~~~~~~~~~~~~~~~~~~~~~~~~~~~~~~~~~~~~~~~~~~~~~~~~~~~~~~~~~~~~
\times j(j-1)(j-2) \cdots (j-m+1) (s-1)^{j-m}, ~~~~~~ |s-1| < 3,
\eqno(78)$$
wherein term-by-term differentiation is justified within the stated radius
of convergence.  Taking $s \to 1^+$ in this equation gives Eq. (29).

For Corollary 6 we write $\eta_{m-1}$ from Eq. (29) and apply the
definition of $S_2$ of Eq. (14) in the form $S_2(n)=\gamma n - \sum_{m=2}^n
{n \choose m} \eta_{m-1}$.  We then apply Lemma 1 for the power series
form of $L_{n-1}^1$ to obtain Eq. (31).

{\em Proposition 5}.  (i) We begin by writing
$${{\zeta'(s)} \over {\zeta(s)}}=-\sum_{n=1}^\infty {{\Lambda(n)} \over n^s}
=\sum_{k=1}^\infty {{\mu(k)} \over k^s} \zeta'(s), ~~~~~~~\mbox{Re} ~s >1,
\eqno(79)$$
and forming $\zeta'(s)$ from Eq. (1).  This gives
$${{\zeta'(s)} \over {\zeta(s)}}=\sum_{k=2}^\infty {{\mu(k)} \over k}
\sum_{\ell=1}^\infty {{(-1)^\ell} \over {\ell !}}\ln^\ell k (s-1)^\ell
\left[-{1 \over {(s-1)^2}}+\sum_{j=0}^\infty {{(-1)^{j+1}} \over {j!}}
\gamma_{j+1} (s-1)^j\right ], \eqno(80)$$
where we have used $\sum_{k=1}^\infty \mu(k)/k=0$.  We then multiply
the series, reorder a double summation, and make use of
$$-\sum_{k=2}^\infty {{\mu(k)} \over k} \ln k = \lim_{s \to 1^+} {d \over {ds}}
{1 \over {\zeta(s)}} = 1,  \eqno(81)$$
wherein we appealed to a Tauberian theorem (\cite{widder}, Ch. V).
After these operations we compare with the defining Laurent expansion (3)
for the $\eta_j$ coefficients, and Proposition 5(i) follows.  

For part (ii) we instead use the identity
$$-{{\zeta'(s)} \over {\zeta(s)}}=\zeta(s) {d \over {ds}} {1 \over {\zeta(s)}}.
\eqno(82)$$
We use the Laurent expansion (3) for the left side, and expansion (1) for $\zeta(s)$ and the Dirichlet series for $1/\zeta(s)$ on the right side.
Expanding in powers of $s-1$ gives
$${1 \over {s-1}}+\sum_{j=0}^\infty \eta_j (s-1)^j=- \left[{1 \over {s-1}}
+ \sum_{n=0}^\infty {{(-1)^n} \over {n!}}\gamma_n (s-1)^n \right]$$ 
$$\times \sum_{\ell=2}^\infty {{\mu(\ell)} \over \ell} \ln \ell
\left[1+\sum_{j=1}^\infty {{(-1)^j} \over {j!}}\ln^j \ell (s-1)^j\right].  \eqno(83)$$
We then expand the right side, reordering the last series, apply relation (81)
and (ii) follows.

Corollary 7 follows from part (i) by using the definition (14) of $S_2(n)$ and applying Lemma 1.  Similarly, another form of $S_2(n)$ could be written based
upon the expression for $\eta_j$ given in part (ii).

{\em Proposition 6}.  The key starting point of the proof is the Amore
representation of the Riemann zeta function \cite{amore}
$$\zeta(s)={1 \over {1-2^{1-s}}} {1 \over {(1+\lambda)}} \sum_{k=0}^\infty
\left({\lambda \over {1+\lambda}}\right)^k \sum_{j=0}^k (-1)^j {k \choose j}
{1 \over \lambda^j} {1 \over {(j+1)^s}}, ~~~~s \in C, ~~~~s \neq 1. 
\eqno(84)$$
Beyond the condition Re $s >0$ stated in Ref. \cite{amore}, the
representation (84) holds for all complex $s \neq 1$, as the summation
continues to converge for Re $s \leq 0$.  (We further discuss Eq. (84) in the
following section.)  So Eq. (84) is globally convergent, as is its $\lambda=1$
special case embodied in the Hasse representation \cite{hasse,coffeyjmaa}.
Moreover, beyond the original condition $\lambda >0$ of Ref. \cite{amore},
we may take $\lambda$ complex with Re $\lambda >0$.

The proof now proceeds as in Proposition 6.1 of Ref. \cite{coffeyprsa}.
However, that description was terse and we now have the (arbitrary)
complex parameter $\lambda$.  Therefore, we believe it is worthwhile to
supply some more details.  We first write again
$$(j+1)^{-s}=(j+1)^{-1}\exp[-\ln (j+1) (s-1)]$$
$$={1 \over {j+1}}\sum_{\ell=0}^\infty {{(-1)^\ell} \over {\ell !}}\ln^\ell (j+1) (s-1)^\ell. \eqno(85)$$
From the generating function of the Bernoulli numbers, we have
$${{te^t} \over {e^t-1}}={t \over {1-e^{-t}}}=\sum_{n=0}^\infty B_n(1){t^n 
\over {n!}}=\sum_{n=0}^\infty (-1)^n B_n {t^n \over {n!}}, ~~~~~~|t| < 2\pi,
\eqno(86)$$
where $B_n(x)$ is the $n$th Bernoulli polynomial.  We then put
$t=(s-1)\ln 2$ in Eq. (86) and obtain
$$(1-2^{1-s})^{-1}={1 \over {\ln 2(s-1)}}-\sum_{j=0}^\infty {{(-1)^j B_{j+1}} \over
{(j+1)!}} \ln^j 2 (s-1)^j, ~~~~~~~~|s-1| < {{2\pi} \over {\ln 2}}, \eqno(87)$$

%Inserting these expressions into Eq. (11) of Ref. \cite{coffeykremin}, rearranging
%sums and separating the polar term using Eq. (13) of Ref. \cite{coffeykremin} gives
%the expression (33).
We substitute Eqs. (85) and (87) into (84), writing the sum over $\ell$ in
Eq. (84) as the $\ell=0$ term and the rest of the terms.  We use
$$\sum_{k=0}^\infty \left({\lambda \over {1+\lambda}}\right)^k \sum_{j=0}^k (-1)^j {k \choose j} {1 \over \lambda^j} {1 \over {(j+1)}}=(\lambda+1)\ln 2, \eqno(88)$$
obtaining 
$$\zeta(s)={1 \over {1+\lambda}}\left[{1 \over {\ln 2(s-1)}}-\sum_{j=0}^\infty {{(-1)^j B_{j+1}} \over {(j+1)!}} \ln^j 2 (s-1)^j\right]$$
$$\times \left[(\lambda+1)\ln 2+ \sum_{k=0}^\infty \left({\lambda \over {1+\lambda}}\right)^k \sum_{j=0}^k (-1)^j {k \choose j} {1 \over \lambda^j} 
{1 \over {(j+1)}}\sum_{\ell=1}^\infty {{(-1)^\ell} \over {\ell !}}\ln^\ell (j+1) (s-1)^\ell \right]. \eqno(89)$$
We multiply the terms in Eq. (89), separate out the simple polar part, and
compare with the defining expansion (1) for the Stieltjes constants.
For the last product of series in Eq. (89) we use the reordering
$$\sum_{n=0}^\infty \sum_{\ell=1}^\infty (\ldots) (s-1)^{n+\ell}
=\sum_{\ell=1}^\infty \sum_{m=\ell}^\infty (\ldots)(s-1)^m=\sum_{m=1}^\infty
\sum_{\ell=1}^m (\ldots) (s-1)^m, \eqno(90)$$
and the Proposition follows.
%later, in the discussion section, cover s=0 special cases, etc. of the
%Amore zeta function repr.

In particular, at $\lambda=1/2$ we obtain
{\newline \bf Corollary 13}.
$$\gamma_m=-{2 \over 3}m!\sum_{\ell=1}^m {{B_{m-\ell+1}} \over {(m-\ell+1)!}}{{\ln^{m-\ell}2} \over {\ell!}}\sum_{k=1}^\infty \left({1 
\over 3}\right)^k \sum_{j=1}^k (-1)^j {k \choose j} 2^j {{\ln^\ell
(j+1)} \over {(j+1)}}$$
$$-{2 \over {3\ln 2}}{1 \over {(m+1)}}\sum_{k=1}^\infty \left({1 \over 3}\right)^k \sum_{j=1}^k (-1)^j {k \choose j} {{\ln^{m+1} (j+1)} \over {(j+1)}}\ln^{m+1} (k+1)
-{B_{m+1} \over {(m+1)}} \ln^{m+1} 2. \eqno(91)$$

Remarks.  The expression (91) may be attractive for some computational
applications because it exhibits even faster convergence than the form
of the Stieltjes constants resulting from the Hasse representation of the
zeta function \cite{coffeyprsa}.  The free parameter $\lambda$ in Eq. (84)
may even be taken as a function of $s$.  Based upon the principle of minimal
sensitivity, one chooses to lowest order $\lambda^{(1)}=2^{-s}$.  
Amore \cite{amore} has noted that with this choice in Eq. (84) a still
exact expression for the zeta function is obtained.  Since in obtaining
the Stieltjes constants we expand about $s=1$ we chose $\lambda=1/2$ in
Corollary 13.  
The form of $\zeta(s)$ given in Eq. (23) of Ref. \cite{amore}, where $s$
dependence now appears in $4$ places, could again be expanded about $s=1$
but this additional effort does not seem warranted. 

Proposition 6 shows once again that the Stieltjes constants may be written
in terms of the Bernoulli numbers and elementary constants such as powers of
the logarithm of the natural numbers.  In turn, the same holds for any
quantity expressible in terms of the Stieltjes constants.  While not many
arithmetic properties are known for the Stieltjes constants, the Bernoulli
numbers have many known important properties such as congruence or other
relations.  For instance, writing the (rational) Bernoulli number $B_j=
\mbox{numer}(B_j)/\mbox{denom}(B_j)$, we have denom$(B_{2n})=\prod_{(p-1)|(2n)}
p$ where the product is over prime numbers $p$ such that $p-1$ divides $2n$.

{\em Proposition 7}.  The Hurwitz zeta function has for Re $s>1$ and Re $a>0$
the integral representation
$$\zeta(s,a)={1 \over {\Gamma(s)}}\int_0^\infty {{x^{s-1} e^{-(a-1)x}} \over 
{e^x-1}} dx={1 \over {\Gamma(s)}}\int_0^1 {{x^{a-1} [\ln(1/x)]^{s-1}} \over
{(1-x)}} dx.  \eqno(92)$$
Then we introduce a parameter $\lambda$ with Re $\lambda>0$, make use of a
geometric series expansion for $x \in [0,1]$, and follow this with a binomial expansion:
$$\zeta(s,a)={1 \over {\Gamma(s)}}\int_0^1 {x^{a-1} \over {(1+\lambda)}}
{{[\ln(1/x)]^{s-1}} \over {\left[1-\left({{x+\lambda} \over {1+\lambda}}
\right)\right]}} dx$$
$$={1 \over {\Gamma(s)}}\int_0^1 {x^{a-1} \over {(1+\lambda)}}[\ln(1/x)]^{s-1}
\sum_{k=0}^\infty \left({{x+\lambda} \over {1+\lambda}}\right)^k dx$$
$$={1 \over {\Gamma(s)}}\sum_{k=0}^\infty {\lambda^k \over {(1+\lambda)^{k+1}}}
\sum_{j=0}^k {k \choose j} {1 \over \lambda^j} \int_0^1 x^{a+j-1} [\ln(1/x)]^{s-1}
dx.  \eqno(93)$$
Performing the integral in terms of the $\Gamma$ function gives the
Proposition.

{\em Proposition 8}.  We first obtain an alternating binomial sum by 
integrating the Beta function.  We have 
$$B(x,y)=\int_0^1 u^{x-1}(1-u)^{y-1}du = 2\int_0^{\pi/2} \sin^{2x-1} \phi
\cos^{2y-1} \phi ~d\phi = B(y,x),$$
$$~~~~~~~~~~~~~~~~~~~~~~~~~~~~~~~~~~~~~~~~~~~~~~~~~~~~~~~~~~~~~~ \mbox{min}[\mbox{Re} ~x, \mbox{Re} ~y]>0, \eqno(94)$$
so that
$$\int_a^t B(x,y)dx= \int_0^1 {{(u^{t-1}-u^{a-1})} \over {\ln u}} (1-u)^{y-1}
du \eqno(95a)$$
$$=\sum_{j=0}^\infty (-1)^j {{y-1} \choose j} \int_a^t {{dx} \over {x+j}} \eqno(95b)$$
$$=\sum_{j=0}^\infty (-1)^j {{y-1} \choose j}[\ln(t+j)-\ln(a+j)],  \eqno(95c)$$
where the form (95b) follows by binomial expansion in Eq. (94).  Upon
comparing Eq. (95a) and (95c) we have
$$\int_0^1 {u^{t-1} \over {\ln u}} (1-u)^{y-1} du=\sum_{j=0}^\infty (-1)^j
{{y-1} \choose j} \ln (t+j) + C, \eqno(96)$$
where $C$ is a constant to be determined.  A simple way to do this is to
put $y=2$ whereupon $\int_0^1 u^{t-1} [(1-u)/(\ln u)] du=\ln t -\ln(1+t)$.
Therefore, $C=0$ and
$$\int_0^1 {u^{t-1} \over {\ln u}} (1-u)^{y-1} du=\sum_{j=0}^\infty (-1)^j
{{y-1} \choose j} \ln (t+j), \eqno(97)$$
a result closely related to tabulated integrals when $y$ is an integer
\cite{grad} (p. 546).  Now if $y=n+1$, $n \geq 1$ an integer, the sum in
Eq. (97) terminates:
$$\int_0^1 {u^{t-1} \over {\ln u}} (1-u)^n du=\sum_{j=0}^n (-1)^j
{n \choose j} \ln (t+j). \eqno(98)$$
We next multiply each side of this equation by $1/(n-m+1)^q$ and sum on
$n$ from $m$ to $\infty$.  We shift the summation index on the left side,
apply the series definition of the polylogarithm function, and the
Proposition follows.  

Remark.  The Proposition may be extended by analytically continuing to
Re $q \geq 1$.  

{\em Proposition 9}. 
By applying binomial expansion to Eq. (40) we have
$$B(z_1,z_2,x,y)=\sum_{j=0}^\infty (-1)^j {{y-1} \choose j} {{[z_2^{x+j}
-z_1^{x+j}]} \over {(x+j)}}.  \eqno(99)$$
We then integrate this expression on $x$ from $a$ to $t$ and compare
with the integrated form of Eq. (40),
$$\int_a^t B(z_1,z_2,x,y)dx = \int_{z_1}^{z_2} {{(u^{t-1} - u^{a-1})} \over 
{\ln u}} (1-u)^{y-1} du, \eqno(100)$$
and the Proposition follows.  Corollary 9 is the special case of $y=1$.
As a further special case we have
{\newline \bf Corollary 14}
$$\int_a^1 B(2,z_2,x,1)dx = \int_2^{z_2} {{(1-u^{a-1})} \over {\ln u}}du,
\eqno(101)$$
showing the close connection with the logarithmic integral
Li$(z) \equiv \int_2^z dt/\ln t$.  

\medskip
%\pagebreak
\centerline{\bf Discussion}
\medskip

Amplifying that the representation (84) holds for all complex $s \neq 1$ we
easily verify that
$$\zeta(0)=-{1 \over {1+\lambda}}\sum_{k=0}^\infty \left({{\lambda-1} \over
{\lambda+1}}\right)^k = -{1 \over 2}.  \eqno(102)$$
%now see 8/14/06 pm notes--
Furthermore, we have
$$\zeta'(s)=-{1 \over {1-2^{1-s}}}\left[2^{1-s}\ln 2 \zeta(s)+ {1 \over {(1+\lambda)}} \sum_{k=0}^\infty  \left({\lambda \over {1+\lambda}}\right)^k \sum_{j=0}^k (-1)^j {k \choose j} {1 \over \lambda^j} {{\ln(1+j)} \over {(j+1)^s}}\right],$$
$$~~~~~~~~~~~~~~~~~~~~~~~~~~~~~~~~~~~~~~~~~~~~~~~~~~~~s \in C, ~~s \neq 1, \eqno(103)$$
so that
$$\zeta'(0)=-\ln 2+ {1 \over {(1+\lambda)}} \sum_{k=0}^\infty  \left({\lambda \over {1+\lambda}}\right)^k \sum_{j=0}^k (-1)^j {k \choose j} {1 \over \lambda^j} \ln(1+j).  \eqno(104)$$
This value is easily shown to be $-(1/2)\ln 2 \pi$ at $\lambda=1$ and otherwise
Eq. (104) shows that the summation term must evaluate to $(1/2)\ln(2/\pi)$.
In fact, we may demonstrate
{\newline \bf Lemma 2}.  For Re $\lambda>0$, Re $x>0$, and Re $y>0$ we have
(a)
$$\sum_{\ell=0}^n \left(-{1 \over \lambda}\right)^\ell {n \choose \ell}
\ln\left({{y+\ell} \over {x+\ell}}\right) = \int_0^1 {{(u^{y-1}-u^{x-1})} \over
{\ln u}} (1-u/\lambda)^n du, ~~~~~~~~ n \geq 0, \eqno(105)$$
and
(b)
$$\sum_{n=0}^\infty \left({\lambda \over {1+\lambda}}\right)^n 
\sum_{\ell=0}^n \left(-{1 \over \lambda}\right)^\ell {n \choose \ell}
\ln\left({{y+\ell} \over {x+\ell}}\right)=(\lambda+1)\int_0^1 {{(u^{y-1}-u^{x-1})} 
\over {\ln u}} {{du} \over {(u+1)}} \eqno(106)$$
$$=(\lambda+1) \sum_{m=0}^\infty (-1)^m \ln \left({{y+m} \over {x+m}}\right)
\eqno(107)$$
$$=(\lambda+1)\left[\ln {{\Gamma\left({{y+1} \over 2}\right)} \over {\Gamma(y/2)}}
-\ln {{\Gamma\left({{x+1} \over 2}\right)} \over {\Gamma(x/2)}} \right].
\eqno(108)$$

For the proof of part (a), we proceed as in Proposition 8, defining the integral
$$I(q,z,a) \equiv \int_0^1 u^{q-1} (1-au)^{z-1} du.  \eqno(109)$$
We then evaluate
$$\int_x^y I(q,z,a)dq = \int_0^1 {{(u^{y-1}-u^{x-1})} \over {\ln u}} (1-au)^{z-1}
du \eqno(110)$$
also by means of binomial expansion of the integrand.  Putting $z-1=n \geq 0$
an integer then gives the first part of the Lemma.  For part (b) we first
use the result of part (a), interchanging summation and integration and
obtaining Eq. (106).   If we expand the integrand factor $1/(1+u)$ of Eq. (106)
as a geometric series, valid for $|u|<1$, and then use a tabulated integral
\cite{grad} (p. 543) we find Eq. (99).  Equation (108) may be found directly
from a known integral \cite{grad} (pp. 543 or 544) applied to Eq. (106), or by
using the Hadamard product representation of the Gamma function in conjunction 
with Eq. (107).  For the latter we note (cf. \cite{grad}, p. 936)
$${{\Gamma\left({{y+1} \over 2}\right)} \over {\Gamma(y/2)}}
{{\Gamma(x/2)} \over {\Gamma\left({{x+1} \over 2}\right)}}
=\prod_{k=0}^\infty \left(1-{1 \over {(2k+y+1)}}\right)\left(1+{1 \over {2k+x}}
\right).   \eqno(111)$$
Therefore we obtain
$$\ln {{\Gamma\left({{y+1} \over 2}\right)} \over {\Gamma(y/2)}}
{{\Gamma(x/2)} \over {\Gamma\left({{x+1} \over 2}\right)}}
=\sum_{k=0}^\infty \ln \left({{2k+y} \over {2k+x}}\right)
\left({{2k+1+x} \over {2k+1+y}}\right)=\sum_{k=0}^\infty \left[
\ln \left({{2k+y} \over {2k+x}}\right)-\ln \left({{2k+1+y} \over {2k+1+x}}\right)
\right]$$
$$=\sum_{m=0}^\infty (-1)^m \ln \left({{m+y} \over {m+x}}\right), \eqno(112)$$
and the Lemma is again completed.  Alternatively, we may directly relate the left
side of (106) to the right side of (107) simply by reordering the double summation
as $\sum_{n=0}^\infty \sum_{\ell=0}^n = \sum_{\ell=0}^\infty \sum_{n=\ell}^\infty$.

Based upon a very special case of Lemma 2(a) we have
{\newline \bf Corollary 15}.  We have (a) for Re $y > -1$
$$\int_0^1 {{(u^y-1)} \over {\ln u}}du = \ln(y+1), \eqno(113)$$
(b)
$$\sum_{j=0}^k (-1)^j {k \choose j} {1 \over \lambda^j} \ln(j+1)
=\int_0^1 \left[\left(1-{u \over k}\right)^k - \left(1-{1 \over \lambda}\right
)^k\right]{{du} \over {\ln u}}, \eqno(114)$$
and (c)
$$\sum_{k=0}^\infty \left({\lambda \over {1+\lambda}}\right)^k \sum_{j=0}^k (-1)^j
{k \choose j} {1 \over \lambda^j} \ln(j+1) = {1 \over 2}(\lambda+1)\ln\left(
{2 \over \pi}\right).  \eqno(115)$$
For part (a), we put $n=0$, $x=1$, and $y \to y+1$ in Eq. (105).  For part (b),
we use the integral representation of part (a) and evaluate the two binomial
sums.  For part (c) we find, using part (b),
$$\sum_{k=0}^\infty \left({\lambda \over {1+\lambda}}\right)^k \sum_{j=0}^k (-1)^j
{k \choose j} {1 \over \lambda^j} \ln(j+1)=\sum_{k=0}^\infty \int_0^\infty
{1 \over {(\lambda+1)^k}}\left[(\lambda-u)^k-(\lambda-1)^k \right] {{du} \over
{\ln u}}$$
$$={{(\lambda+1)} \over 2} \int_0^1 \left({{1-u} \over {1+u}}\right) {{du} \over {\ln u}} = {1 \over 2}(\lambda+1)\ln \left({2 \over \pi}\right).  \eqno(116)$$
Of many ways to evaluate the last integral of Eq. (116), one may use \cite{grad}
(p. 542).  The Corollary is demonstrated and Eq. (104) is affirmed.

Moreover, we must recover the values $\zeta(-n)=(-1)^n B_{n+1}/(n+1)$ for $n$
an integer from Eq. (84).  For $n$ even this includes the trivial zeros of the zeta function, whereby $B_{2m+1}=0$ for $m \geq 1$.  We have from Eq. (84)
$$\zeta(-n)={1 \over {(1-2^{n+1})}} {1 \over {(1+\lambda)}} \sum_{k=0}^\infty \left({\lambda \over {1+\lambda}}\right)^k \sum_{j=0}^k (-1)^j {k \choose j}
{1 \over \lambda^j} (j+1)^n.  \eqno(117)$$
For $\lambda=1$ this equation recovers the old formula of Worpitsky for the 
Bernoulli numbers \cite{carlitz,worp}.

As a byproduct of this work we obtain interesting infinite series (or products)
for fundamental constants such as the Euler constant and $\ln \pi$.  The
rapidity of convergence may make some of these suitable for computational applications.  We omit many such binomial summations that may be obtained by
methods very close to Propositions 8 and 9.
%Indeed, even a naive numerical implementation of part (a)
%or (b) of Proposition 8 appears to provide these constants to 16 decimal
%places after summing over $n$ to $51$ or $52$ terms.

%Later--on the various integrals and alt. binomial sums from integrating the
%Beta func., correspondence with some G/R formulae, etc. ............... TO DO ...

With respect to the right side of Eq. (12), integers $m$ that are powers of
$2$ play a special role.  It is only these contributions for which 
$1-\Lambda(m) =1-\ln 2>0$, while all other integral powers of primes 
give $1-\Lambda(m)<0$.  %may need to reword a bit in here ....

%Comments on relation of $S_2(n)$ to the Li criterion for the RH ...
%it all depends upon the behaviour of $S_\Lambda$ ........................
Within the Li criterion for the Riemann hypothesis \cite{li97}, the sum 
$S_2(n)$ is given by (e.g., \cite{coffeympag})
$$S_2(n)=\sum_{m=1}^n {n \choose m} {1 \over {(m-1)!}}\left({d \over {ds}}
\right)^m \left[\ln(s-1)\zeta(s)\right]_{s=1}.  \eqno(118)$$
Therefore, by using the Hadamard product representation of the Riemann
zeta function, it is easy to see that $S_2(n)$ is connected with sums over
its nontrivial zeros.  If $S_2(n)$ has linear or sublinear growth in $n$, the
Riemann hypothesis holds.  According to our decomposition (15), the
validity of the Riemann hypothesis is equivalent to the sum $S_\Lambda(n)$ 
having linear or sublinear growth in $n$.  In fact, we conjecture (see below)
that $|S_\Lambda(n)| =O(n^{1/2+\epsilon})$ for $\epsilon >0$ arbitrary.
(This conjecture is stronger than the Riemann hypothesis itself.)

In regard to Proposition 4, we put 
$$\Delta_2(x) \equiv \sum_{n \leq x} d(n) -x \ln x -(2\gamma-1)x=O(x^{\alpha_2+
\epsilon}), ~~~~~~~~\epsilon > 0,  \eqno(119)$$
wherein $\alpha_2$ is the least such number for every positive $\epsilon$.
Dirichlet knew that $\alpha_2 \leq 1/2$ and the best result to date may
be $\alpha_2 \leq 131/416$ \cite{huxley3}.  In fact, Huxley showed that
$\Delta_2(x) = O(x^{23/73} \ln^{461/146} x)$ \cite{huxley2} and improved 
this very recently to $\Delta_2(x) = O(x^{131/416})$ \cite{huxley3}.   
%$\alpha_2 \leq 139/429$ \cite {kolesnik}.  
The smallest possible value of
$\alpha_2$ is $1/4$, and we prove that $|S_\gamma(n)+n| = O(n^{1/4})$
(see after Eq. (120) below).
The method of Propositions 1 and 4 is easily extended to many other pole-free
combinations of Dirichlet series.

Proposition 6, its special case Proposition 6.1 of Ref. \cite{coffeyjmaa},
or other series representations of the Stieltjes constants may be used
to obtain alternative summation representations of the $O(n)$ sum $S_\gamma(n)$.

%add figure for $S_\gamma(n)+n$ and $S_\Lambda(n)-n$ later? .................
Numerical investigations indicate that $S_\gamma(n)$ is close
to $-n$ together with a small oscillatory component, while $S_\Lambda(n)$ is
close to $n$ with a small oscillatory component $S_{2\Lambda}(n)$
\cite{cofexpmath}.  Therefore, 
the crucial sum $S_2(n)$ appears to arise from substantial cancellation of
$O(n)$, leaving a slowly growing, oscillatory contribution.   
%more work in progress on this ...
%follow up on trying to break out m=2 term in S_\Lambda(n), etc.  see my notes
%and do more on this in the future ......
A demonstration that $S_2(n)$ satisfies a one-sided subexponential bound
would suffice to verify the Riemann hypothesis. 

As a point of emphasis, the Riemann hypothesis will only fail if a Li/Keiper
constant $\lambda_k$ becomes exponentially large in magnitude and negative.
In particular, the Criterion (c) of Ref. \cite{bombieri} now carries over to
the crucial subsum $S_{2\Lambda}(n)$.  Therefore the Riemann hypothesis
is invalid only if this sum becomes negative and exponentially large in 
magnitude for some $n$.  %(We know that such $n$ has to exceed $10^{26}$ and
%would have to be probably much larger.)
We may spell this out in the following way.
{\newline \bf Condition}.  Suppose that there is a value of $1/2 \leq p <
\infty$ such that $|S_{2\Lambda}(n)| = O(n^p)$.

Then upon Condition the Riemann hypothesis will follow as a Corollary.  It
is compelling within the Li criterion approach that the optimal order of the
sum $S_{2\Lambda}(n)$ is not necessarily required.  As indicated, we suspect
that the lowest possible order of this sum is close to $O(n^{1/2})$.

From Fej\'{e}r's formula for the asymptotic form of $L_n^\alpha(x)$ we have
for $n \to \infty$
$$L_{n-1}^1(x)={1 \over \sqrt{\pi}}e^{x/2} x^{-3/4} (n-1)^{1/4} \cos(2\sqrt{(n-1)x}
-3\pi/4) + O(n^{-1/4}), ~~~~~~~~ x > 0.  \eqno(120)$$
Therefore we now show that the oscillatory component of $S_\gamma(n)+n$ 
grows as $O(n^{1/4})$.  Indeed, the last two terms on the right side of
Proposition 2(d) contribute at $O(n^{1/4})$, with $\cos(2\sqrt{n-1}y+\phi)$
factors.  The last three terms on the right side additionally contribute
at the next lowest order of $n^{-1/4}$.  For the remaining term on the
right side of Proposition 2(d) we have
$$-\sum_{\nu=1}^N {{L_{n-1}^1(\ln \nu)} \over \nu}=-n
+\sum_{\nu=2}^N {{L_{n-1}^1(\ln \nu)} \over \nu}=-n+O(N^{1/2+\delta}n^{1/4}),
\eqno(121)$$
with $\delta>0$.  We may therefore summarize as
{\newline \bf Corollary 16}.  As $n \to \infty$ we have
$|S_\gamma(n)+n| = O(n^{1/4})$.

Besides the indications given in Ref. \cite{coffeympag} that the Laguerre
calculus is pervasive within the Li/Keiper formulation of the Riemann
hypothesis, we have very recently systematically presented the structural
origins of this framework \cite{coffeyjat}.  %just what other property(ies)
%single these polys out? -- divisibility properties or what ?? TBD ...
The Li/Keiper constants arise as a sum over complex zeta zeros of a
Laplace transform of the associated Laguerre polynomial $L_{n-1}^1(x)$.
We have  
$$L_n(\rho) \equiv \int_0^\infty e^{-\rho u}L_{n-1}^1(u)du=1-\left(1-{1 \over \rho}\right)^n, \eqno(122)$$
that vanishes for $\rho_j=(1-e^{2\pi i j})^{-1}$, with $j=1,\ldots,n-1$.
These Laplace transform zeros have real part $1/2$:
$$\rho_j={1 \over 2}{{(1-e^{-2\pi ij/n})} \over {[1-\cos(2\pi j/n)]}}
={1 \over 2}\left[1+ i \cot\left({{\pi j} \over n}\right)\right]. \eqno(123)$$
That these quantities lie on the critical line may be more than just a
curiosity and may partly explain the distinguished role of the
polynomial $L_{n-1}^1$.

Furthermore, we recall that \cite{coffeympag} (Appendix I) 
$$L_n(s)=\int_0^1 x^{s-1}L_{n-1}^1(-\ln x)dx = {n \over s} ~_2F_1\left(1-n,1;2;
{1 \over s}\right), \eqno(124)$$
in terms of the Gauss hypergeometric function.  We put 
${\cal L}_n(s)=s^n L_n(s)$ and have
{\newline \bf Corollary 17}.  We have the functional equation
$$L_n(s)=-\left(1-{1 \over s}\right)^n L_n(1-s), \eqno(125)$$
and
$${\cal L}_n(s)=(-1)^{n+1} {\cal L}_n(1-s),  \eqno(126)$$
that follows immediately.  Equation (125) may be obtained directly or by
applying the transformation formula \cite{grad}
$$_2F_1(\alpha,\beta;\gamma;z)=(1-z)^{-\alpha} ~_2F_1\left (\alpha,\gamma-\beta;\gamma;{z \over {z-1}}\right ) \eqno(127)$$
to the right side of Eq. (124).
%so now cite my PLA to appear?

Propositions 1--5 and other results that we have obtained help to expose
more of the analytic structure of the Stieltjes and $\eta_j$ constants. 
We have obtained novel integral and other representations of the Stieltjes
constants that enable new integral and series representations of the sums $S_\gamma(n)$ and $S_2(n)$.  The growth behavior of $S_{2\Lambda}(n)$ and
$S_2(n)$ have direct implication for the validity of the Riemann hypothesis.

\bigskip
\centerline{\bf Acknowledgement}
\medskip
I thank R. Kreminski for access to high-precision values of $\gamma_k$. 

%consider if to later add two (three) figures say for ($S_2(n)$?), 
%$S_\gamma(n)+n$, and $S_\Lambda(n)-n$ ...................

%\pagebreak
%\medskip
%\centerline{\bf Figure Caption}

%FIG. 1.  ...

\pagebreak

\end{document}